\documentclass[useAMS,usenatbib,usegraphicx]{mn2e}

\newcommand{\kev}{keV}

\newcommand{\fe}{Fe~K$\alpha$}
\newcommand{\etal}{et al.}
\newcommand{\mcg}{MCG--6-30-15}

\title[Ionized accretion disc models of MCG--6-30-15]
  {Relativistic ionized accretion disc models of MCG--6-30-15}
\author[D.\ R.\ Ballantyne \& A.\ C.\ Fabian]
  {D.~R.~Ballantyne\thanks{drb@ast.cam.ac.uk} and A.~C.~Fabian\\
  Institute of Astronomy, Madingley Road, Cambridge CB3 0HA}

\pagerange{\pageref{firstpage}--\pageref{lastpage}}
\pubyear{2001}

\usepackage{times}

\begin{document}

\label{firstpage}

\maketitle

\begin{abstract}
We present results from fitting ionized accretion disc models to three
long \textit{ASCA} observations of the Seyfert 1 galaxy \mcg. All
three datasets can be fit by a model consisting of ionized reflection
from the inner region of the accretion disc (with twice solar Fe
abundance) and a separate diskline component from farther out on the
the disc. The diskline is required to fit the height of the observed
\fe\ line profile. However, we show that a much simpler model of
reflection from a very weakly ionized constant density disc also fits
the data. In this case only a single cold \fe\ line at 6.4~\kev\ is
required to fit the observed line. The ionized disc models predict
that O~\textsc{viii} K$\alpha$, C~\textsc{vi} K$\alpha$,
Fe~\textsc{xvii} L$\alpha$, and Fe~\textsc{xviii} L$\alpha$ lines will
appear in the soft X-ray region of the reflection spectrum, but are
greatly blurred due to Compton scattering. The equivalent width (EW)
of O~\textsc{viii} K$\alpha$ is estimated to be about 10~eV and seems
to be as strong as the blend of the Fe L lines. This result creates
difficulty for the claim of a strong relativistic O~\textsc{viii} line
in the \textit{XMM-Newton} grating spectrum of \mcg, although we can
not strictly rule it out since \mcg\ was in an anomalously low state
during that observation. We find that increasing the O abundance or
breaking the continuum below 2~\kev\ will not significantly strengthen
the line. The second \fe\ line component in the ionized disc model may
arise from neutral reflection from a flared disc, or from a second
illumination event. The data cannot distinguish between the two cases,
and we conclude that single zone ionized disc models have difficulty
fitting these hard X-ray data of \mcg.
\end{abstract}


\section{Introduction}
\label{sect:intro}
\mcg\ is a bright nearby ($z=0.008$) Seyfert 1 galaxy, and has thus
been observed by various X-ray telescopes over the last 20 years
\citep*[e.g.,][]{pin80,pou86,nan89,nan90,fab94}. These studies
culminated in the first detection of a relativistically broadened iron
K$\alpha$ line \citep{tan95}. The shape of the line is consistent with
being emitted from fluorescing, optically thick material which is
orbiting very close to a black hole \citep{fab89}, and is difficult to
explain by other processes \citep{fab95}. Subsequent deep
\textit{ASCA} observations have found evidence for broad \fe\ lines in
other Seyfert 1 galaxies \citep{n97}, illustrating that these features
may be common in type 1 Active Galactic Nuclei (AGN). Due to the
proximity of \mcg, its \fe\ line has been studied many times at high
signal-to-noise and has been shown to be variable, both in strength
and in shape \citep*{iwa96,iwa99,lee00,vau01}.

Recently, \citet[][hereafter BR01]{bra01} reported results from
observations of \mcg\ and another Seyfert 1 galaxy Mrk~766 using the
Reflection Grating Spectrometer (RGS) on \textit{XMM-Newton}. These
authors claim that features in the soft X-ray band that were
traditionally interpreted in lower-resolution data as photoelectric
absorption edges could actually be relativistically broadened
K$\alpha$ lines of O~\textsc{viii}, N~\textsc{vii} and
C~\textsc{vi}. However, a non-simultaneous observation of \mcg\ by the
High-Energy Transmission Grating (HETG) on \textit{Chandra} (which has
higher spectral resolution than the RGS) does not confirm the above
interpretation \citep{lee01}. These authors find that the spectral
features in the soft X-rays are well fit by a dusty warm absorber
model.

Regardless of which interpretation is correct, the idea of other
relativistic emission lines in the X-ray spectra of Seyfert galaxies
is intriguing, and is worth further investigation. This can be done by
employing ionized reflection models
\citep*[e.g.,][]{ros93,ros99,nkk00,bal01} to predict the reflection
spectrum from an ionized disc. These models cannot be fit to grating
data as there is no \fe\ line or easily determined continuum to
``anchor'' the model. Therefore, in this paper, we pursue a
complementary approach, and fit the hard X-ray spectrum of \mcg\ from
\textit{ASCA}, thereby taking advantage of the well-defined \fe\ line
and continuum to fix the parameters of the model. Although the model
is fit only to the high energy data it will have specific predictions
for the strength of the soft X-ray spectral features, which may be
useful in the interpretation of the grating data.

The paper is organized as follows. First, in Sect.~\ref{sect:fits}, we
describe the ionized disc model that is used and present our fits to
the \textit{ASCA} data of \mcg. Then we discuss the strength of the
predicted low-energy features and other consequences of our model in
Sect.~\ref{sect:disc}. Finally, conclusions are drawn in
Sect.~\ref{sect:concl}.

\section{Ionized disc models and fitting}
\label{sect:fits}
Long observations of \mcg\ were obtained by \textit{ASCA} in 1994
(200~ks), 1997 (200~ks) and 1999 (400~ks). A description of the data
reduction can be found in the paper by \citet{iwa96} for the 1994 data
and in the paper by \citet{iwa99} for the 1997 data. Analogous
procedures were used in the reduction of the unpublished 1999 data,
except for the inclusion of a new CTI file
(\texttt{sisph2pi\_130201.fits}) to deal with recent calibration
problems in the SIS detectors. Fits were performed using the
time-averaged data from all four detectors between 3 and
10~\kev. Since the \textit{ASCA} detectors had yet to suffer much
damage, and the calibration is now well established, the experimental
fitting was done using the 1994 data. The model that best fit this
dataset was then used on the 1997 and 1999 dataset to check the
robustness of the model and to investigate variations in the
parameters.

The ionized reflection models were computed with the code described by
\citet{bal01}. These simulations compute the reflection spectrum as
well as the temperature, ionization and density structure of the top
five Thomson depths of an accretion disc that is irradiated by a
power-law continuum of X-rays. As shown by \citet{bal01}, the features
in the reflection spectrum depend on the system parameters (such as
black hole mass, accretion rate and radius along the disc), on the
value of the irradiating flux and photon index of the incident
spectrum ($\Gamma$), and the incidence angle of the radiation
($\theta$). The reflection spectra that were fit to the \textit{ASCA}
data of \mcg\ were computed assuming a black hole mass of
10$^7$~M$_{\odot}$, an accretion rate of 0.01 times the Eddington rate
(so the disc was radiation pressure dominated), and that the
reflection was occurring at 4 Schwarszchild radii from the black
hole. These values are appropriate for a typical Seyfert~1 galaxy like
\mcg. The choice of irradiating flux and incidence angle should
ideally be determined by a specific model for how an accretion disc is
illuminated. We have chosen an illuminating flux to disc flux ratio of
10 (as might be appropriate for irradiation by magnetic flares, e.g.,
\citealt{nk01}), and an incidence angle of $\theta=54.7$
degrees to the normal (so that $\cos \theta=1/\sqrt{3}$, a crude
approximation to isotropic illumination) was assumed.

With these parameters fixed, two sets of reflection spectra were
calculated with $\Gamma$ varied between 1.5 and 2.15: one with solar
abundance of Fe (as given by \citealt{mcm83}) and the other with twice
solar Fe abundance. Each set was weighted by the reflection fraction
$R$ ($0.0 \leq R \leq 3.0$) and then added to the illuminating
power-law. Finally, both 2-dimensional grids were converted into a
\textsc{xspec} `atable' file for use in fitting the data. Relativistic
blurring appropriate for a Schwarszchild metric and assuming a disc
emissivity law (i.e., emissivity $\propto (1-\sqrt{6/r})/r^3$;
\citealt{fab89}) was applied to the spectra during
fitting. The \textsc{xspec} command `extend' was used to increase the
energy range of the response matrices because this convolution
requires evaluating the model at energies outside the 3--10~\kev\
range. Galactic absorption was also included in the fit and fixed at
the value of $4.06 \times 10^{20}$~cm$^{-2}$, but will have a
negligible affect at energies greater than
3~\kev. \textsc{xspec}~v.11.0.1z1 \citep{arn96} was used for fitting,
and the uncertainties in the model parameters are the 1-$\sigma$
errorbars for one interesting parameter.

\subsection{Results of spectral fitting}
\label{sub:res}
The results of the spectral fitting are shown in
Table~\ref{table:fits1}.
\begin{table*}
\begin{minipage}{149mm}
\caption{Results of fitting ionized disc models to the \textit{ASCA}
data of \mcg. The parameters for the diskline are denoted with an $l$
superscript. All radii are in units of the gravitational radius, the
inclination angle $i$ is tabulated in degrees, and the energy of the
diskline $E^l$ is in \kev\ and is reported in the rest-frame of \mcg.}
\label{table:fits1}
\begin{tabular}{@{}cccccccccc}
Model & $\Gamma$ & $R$ & $r_{in}$ & $r_{out}$ & $i$ & $E^{l}$ &
$r^{l}_{in}$ & $r^{l}_{out}$ & $\chi^2$/dof \\ \hline 
1994 & & & & & & & & & \\ \hline 
power-law$^a$ & 1.96$\pm 0.03$ & & & & & & & & 469/496\\ 
1xFe ion. disc & 1.90 & 1.0$^f$ & 15 & 19 & 0.005 & & & & 1754/1653 \\ 
 & 1.92 & 2.88$^{+0.12p}_{-0.63}$ & 16 & 18 & 0.0 & & & & 1684/1652 \\ 
2xFe ion. disc & 1.90 & 1.0$^f$ & 8.8 & 23 & 0.002 & & & & 1722/1653 \\ 
 & 1.91 & 2.06$^{+0.26}_{-0.49}$ & 6.0 & 34 & 0.001 & & & & 1690/1652 \\ 
1xFe ion. disc + diskline & 1.91 & 1.0$^f$ & 6.0 & 11 & 25 & 6.4$^f$ & 21 & 997 & 1653/1650 \\ 
 & 1.92 & 2.27$^{+0.73p}_{-0.98}$ & 6.1 & 10 & 26 & 6.4$^f$ & 61 & 1000 & 1640/1649 \\ 
 & 1.92 & 1.0$^f$ & 6.8 & 10 & 10 & 6.78$^{+0.07}_{-0.18}$ & 15 & 73 & 1633/1649 \\ 
2xFe ion. disc + diskline & 1.91 & 1.0$^f$ & 7.6 & 8.8 & 25 & 6.4$^f$ & 35 & 1000 & 1643/1650 \\ 
 & 1.92 & 1.33$^{+0.64}_{-0.32}$ & 8.0 & 8.4 & 25 & 6.4$^f$ & 53 & 968 & 1640/1649 \\ 
 & 1.91$\pm 0.02$ & 1.0$^f$ & 8.35 & 8.51 & 19$\pm 2$ & 6.55$^{+0.07}_{-0.05}$ & 41 & 533 & 1630/1649 \\
 & broken$^b$ & 1.0$^f$ & 8.36 & 8.51 & 19 & 6.53 & 35 & 151 & 1679/1650 \\ \hline
1997 & & & & & & & & & \\ \hline 
power-law$^a$ & 1.93$\pm 0.03$ & & & & & & & & 483/488 \\ 
2xFe ion. disc + diskline & 1.89$^{+0.02}_{-0.03}$ & 1.0$^f$ & 6.4 & 7.0 & 23 & 6.4$^f$ & 35 & 971 & 1694/1619 \\ 
 & 1.90 & 1.89$^{+0.48}_{-0.77}$ & 6.1 & 7.1 & 26.9 & 6.4$^f$ & 55 & 998 & 1689/1618 \\
 & 1.89 & 1.0$^f$ & 6.6 & 7.0 & 20 & 6.53 & 30 & 154 & 1684/1618 \\ \hline 
1999 & & & & & & & & & \\ \hline
power-law$^a$ & 2.02$\pm 0.02$ & & & & & & & & 727/673 \\ 
2xFe ion. disc + diskline & 1.96$^{+0.02}_{-0.01}$ & 1.0$^f$ & 6.9 & 7.1 & 26 & 6.4$^f$ & 20 & 999 & 1924/1893 \\
 & 1.97 & 1.94$^{+0.52}_{-0.32}$ & 6.0 & 7.9 & 28 & 6.4$^f$ & 56 & 485 & 1893/1892 \\
 & 1.96 & 1.0$^f$ & 6.0 & 7.8 & 19 & 6.66$\pm 0.06$ & 25 & 80 & 1868/1892 \\ \hline
\end{tabular}

\medskip
$^a$ 4--8~\kev\ data not included in fit

$^b$ $\Gamma=1.91$ for $E > 2.0$~\kev, $\Gamma=1.33$ for $E < 2.0$~\kev.

$^f$ Parameter fixed at value

$^p$ Parameter pegged at upper limit
\end{minipage}
\end{table*}
The first model tabulated is a simple absorbed power-law model that
was fit to the data in the range 3--4~\kev\ and 8--10~\kev. This fit
gives a measurement of the underlying continuum in the data, and we
would like the more complex ionized disc models to give a photon index
that is compatible with this value.

Fitting the data with a straightforward ionized disc model does not
give a very satisfactory result. The line in the model is not strong enough
to fit the data, and the fit can only be improved by increasing the
reflection fraction, $R$. Furthermore, the derived photon-index is not
consistent with the power-law fit, and the inclination angle is very small.

A more satisfying, although still problematic, fit can be obtained by
doubling the Fe abundance in the disc model. That \mcg\ might have an
overabundance of Fe was first suggested by \citet{lee99} in order to
explain the strength of the \fe\ line seen in the 1997 \textit{ASCA}
spectrum. These authors also utilize higher energy data from
\textit{RXTE} to constrain the reflection fraction to be around
one. Indeed, the twice solar Fe model lowers the reflection fraction,
but the model still cannot account for all the flux in the \fe\
line. The problem is that the predicted \fe\ lines from the ionized
disc models are not strong enough to simultaneously fit both the large
red wing and the large peak of the line profile, and thus the Fe
abundance in \mcg\ would have to be very large to account for the \fe\
line. A simple and straightforward way to resolve this problem is to
add in another line component \citep[cf.,][]{mis01}.
\begin{figure}
\includegraphics[angle=-90,width=0.50\textwidth]{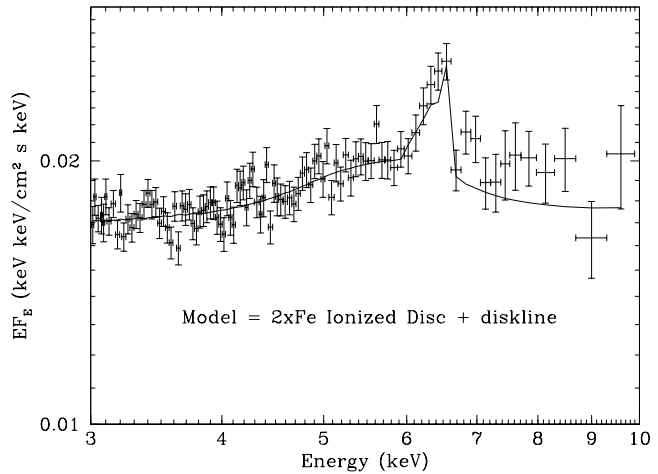}
\caption{The unfolded 1994 \textit{ASCA} SIS spectra of \mcg\ are
plotted along with the best fit ionized disc (with 2 times solar Fe)
and diskline model ($\chi^2/$dof=1630/1649). The ionized disc fit
parameters are: $\Gamma=1.91$, $R=1.0$ (fixed), $r_{in}=8.4$~$r_g$,
$r_{out}=8.5$~$r_g$, and $i=18.7$ degrees. For the diskline model
component the fit parameters are: $E=6.55$~\kev, $r_{in}=41$~$r_g$,
and $r_{out}=533$~$r_g$. The flux in the diskline and ionized line are
$5 \times 10^{-5}$~photons~cm$^{-2}$~s$^{-1}$ and $1.5\times
10^{-4}$~photons~cm$^{-2}$~s$^{-1}$, respectively.}
\label{fig:unfolded}
\end{figure}

The idea that multiple \fe\ components may be common in the line
profiles of Seyfert 1 galaxies has been gaining momentum with the
recent detections of narrow 6.4~\kev\ lines in NGC~3783 \citep{kas01}
and NGC~5548 \citep{yaq01}. However, in this case, a narrow line was
not sufficient to account for the missing line flux, so a broadened
diskline (also calculated with a disc emissivity profile) was added to
the model. With this addition the $\chi^2$ dropped by $\sim$ 80 for
the addition of 3 degrees of freedom -- clearly a significant
improvement in the fit -- and the inclination angle (which was fixed
to be the same for the ionized disc and diskline components) was now
comparable with the results of \citet{tan95} and \citet{iwa99}. With
the addition of the diskline, the iron-rich model now has a reflection
fraction that is consistent with unity and the photon-index is
consistent with the one from the power-law fit. The centroid energy of
the diskline suggests that it arises from partially ionized Fe
($\Delta \chi^2$=$-13$ from $E^{l}$=6.4~\kev) and with originating
from a distance greater than 40~$r_g$ (where $r_g=GM/c^2$, is the
gravitational radius) from the black hole
(Figure~\ref{fig:unfolded}). The relativistic ionized line has three
times the flux of the diskline component: $1.5\times
10^{-4}$~photons~cm$^{-2}$~s$^{-1}$ as compared to $5 \times
10^{-5}$~photons~cm$^{-2}$~s$^{-1}$.

Taking this model to be our best-fit, we then apply it to the 1997 and
1999 datasets of \mcg, and find a good fit in both
cases. Interestingly, the lowest $\chi^2$ are found when the diskline
component arises from partially ionized iron.

\begin{figure}
\includegraphics[angle=-90,width=0.50\textwidth]{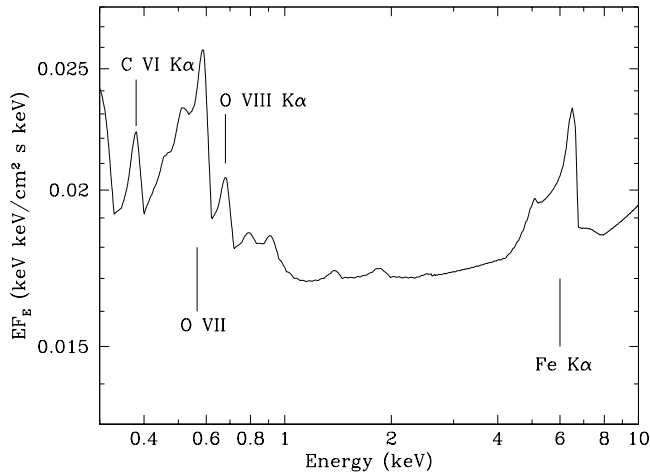}
\caption{The reflection spectrum between 0.3 and 10~\kev\ predicted by
the constant density models of \citet{ros93} after fitting to the 1994
\textit{ASCA} spectrum of \mcg\ ($\chi^2/$dof=1659/1652). The incident
power-law is included in this prediction. The fit parameters are
$\Gamma=1.98\pm0.01$, $\xi$=10--14, $r_{in}=7.5$~r$_g$, $r_{out}=13$~r$_g$, and
$i=28$ degrees. Twice solar Fe abundance was assumed in the model.}
\label{fig:cdens}
\end{figure}

These ionized disc models are admittedly very model dependent, so we
also fit the 1994 \textit{ASCA} data of \mcg\ with the constant
density reflection models of \citet{ros99} which do not require an
assumed black hole mass, accretion rate, etc. After taking the Fe
abundance to be twice solar and a reflection fraction of unity, there
are only three fit parameters: the normalization of the model, the
photon index, and the ionization parameter $\xi=4 \pi F_x/n_{\mathrm
H}$, where $F_x$ is the illuminating flux and $n_{\mathrm H}$ is the
hydrogen number density of the slab. As before, models where the slab
was ionized were unable to fit the \fe\ without a second
component. However, we were able to find a good fit
($\chi^2$/dof=1659/1653) with a $\xi \approx 10$ reflector (see
Fig.~\ref{fig:cdens}). 
This model is practically equivalent to dropping the illuminating flux
to disc flux ratio in the hydrostatic models by a factor $\sim$
10$^3$. Therefore, the easiest fit to the 3--10~\kev\ 1994
\textit{ASCA} data of \mcg\ is with a very weakly illuminated
accretion disc.

\section{Discussion}
\label{sect:disc}

\subsection{Strength of the soft X-ray features}
\label{sub:oxy}
\begin{figure}
\includegraphics[angle=-90,width=0.50\textwidth]{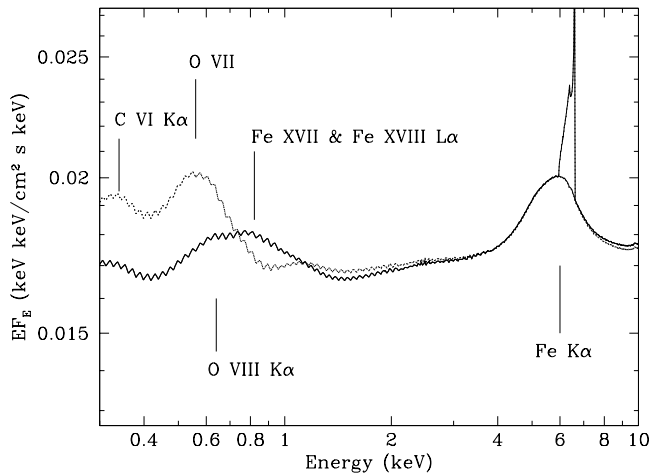}
\caption{The reflection spectrum predicted by the 2$\times$Fe ionized
disc + diskline model when fit to the 1994 \textit{ASCA} data of
\mcg. The incident power-law is included in this prediction. The solid
line denotes the model with solar abundance of oxygen, while the
dotted line shows the model (with the same fit parameters) with five
times solar abundance of oxygen. Emission lines that are emitted by
the disc are indicated in the plot. The high-frequency ``bumpiness''
in the continuum is a result of the interpolation and binning that
\textsc{xspec} must do to produce the figure. Recall that nitrogen is
not included in the model.}
\label{fig:model-pred}
\end{figure}

\begin{figure}
\includegraphics[angle=-90,width=0.50\textwidth]{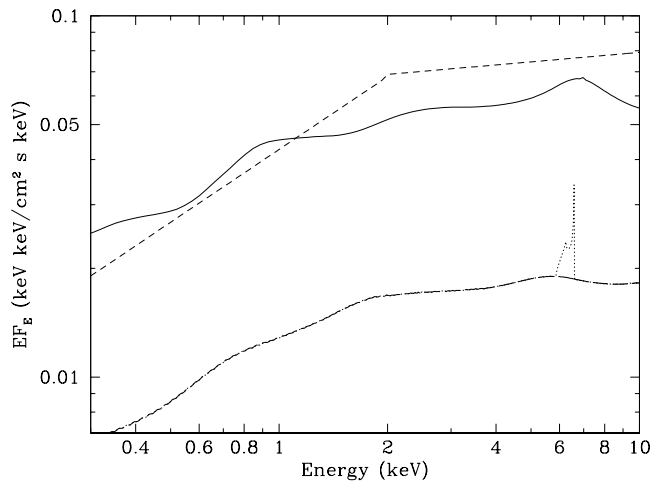}
\caption{The solid line shows the predicted reflection spectrum
from an atmosphere that is illuminated by a broken power-law
continuum ($\Gamma=1.91$ for $E > 2$~\kev\ and $\Gamma=1.33$ for $E <
2$~\kev; short-dashed line). Both of these curves have been scaled
downwards by a factor of $5\times10^{25}$. When the reflection spectrum is
added to the continuum and fit to the 1994 \textit{ASCA}
data of \mcg\, the resulting best-fit ($\chi^2$/dof=1679/1650) is
given by the long-dashed and dotted lines. The resulting
O~\textsc{viii} line is very weak.}
\label{fig:model-broken}
\end{figure}

Figure~\ref{fig:model-pred} shows the prediction of the best fit
ionized disc model to the 1994 \textit{ASCA} data of \mcg. Recall that
the reflection fraction is one, so the illuminating power-law is
included.
The model predicts highly ionized features that have suffered
significant Compton broadening due to the ionized gas at the surface
of the illuminated atmosphere. These features include the K$\alpha$
lines of C~\textsc{vi} and O~\textsc{viii}, and a blended component
comprised of L$\alpha$ lines of Fe~\textsc{xvii} and
Fe~\textsc{xviii}. Nitrogen is not included in our calculations, so we
are unable to place constraints on its strength. As a result of the
Compton broadening, it was difficult to accurately measure the
equivalent widths (EW) of the O line and the Fe L blend. The EW
estimate was made by assuming that the O line was equally as strong as
the Fe L blend. The weakly ionized constant density model does not
suffer from much Comptonization, and so the EWs were calculated
directly from the spectrum. All the EW results are shown in
Table~\ref{table:ews}.
\begin{table}
\caption{Equivalent widths in eV of the soft X-ray features predicted
by ionized disc models of \mcg.}
\label{table:ews}
\begin{center}
\begin{tabular}{@{}cccc}
O abundance & C~\textsc{vi} & O~\textsc{viii}\\ \hline
solar & 5.0 & 6$^a$\\
 & 4.2$^\dag$ & 5.4$^\dag$\\ 
3$\times$ solar & 7.7 & 16$^a$\\
5$\times$ solar & 7.3 & 12$^a$\\ \hline
\end{tabular}

\medskip
$^a$ estimated as half the EW of the O~\textsc{viii}-Fe L blend

$^\dag$ from the constant density model (Fig.~\ref{fig:cdens})
\end{center}
\end{table} 
We find that the O line has an EW about 6~eV. This is much smaller
than the EW$\sim$150~eV O~\textsc{viii} K$\alpha$ line claimed by BR01
using their RGS observation of \mcg. However, the comparison is not
strictly valid, as BR01 requires that the continuum turn over so that
$\Gamma=1.33$ below $\sim$ 2~\kev. To check how this break in the
continuum will affect the strength of the soft X-ray features, we
computed a model with $\Gamma=1.91$ for $E > 2$~\kev\ and
$\Gamma=1.33$ for $E < 2$~\kev, and fitted this to the 1994 data
(Figure~\ref{fig:model-broken}).
The calculation shows that the hard incident spectrum at low energies
ionizes the disc even more and therefore further weakens the emission
features. The EW of the O~\textsc{viii} line is estimated to be $\sim
1$~eV.

Returning to models with unbroken power-laws, we investigated
abundance effects by running computations with three times and five
times solar oxygen abundances. Substituting these models for the solar
abundance model did not greatly affect the $\chi^2$ of the fit. The
EWs of the soft lines in these models are also shown in
Table~\ref{table:ews}. When the O abundance is increased to 5 times
solar, O~\textsc{vii} recombination lines are prominent and are
blended with the O~\textsc{viii} line (see
Fig.~\ref{fig:model-pred}). The resolution of the RGS is high enough
that lines from O~\textsc{vii} should be distinguishable from
O~\textsc{viii} K$\alpha$ even with extreme relativistic blurring, so
a supersolar abundance of O cannot account for the strength of the
line claimed by BR01. Finally, it is important to note that if there
is another soft component in the spectrum, such as a soft excess, then
the EWs will be even smaller than those measured here.

Another important prediction of these ionized disc models is that
there should be L$\alpha$ lines from Fe~\textsc{xvii} and
Fe~\textsc{xviii} which, when blended together, seem to have about the
same strength as the O~\textsc{viii} K$\alpha$ line. These are not
seen\footnote{We note in passing that there is a slight, unexplained
excess at $\sim$ 0.8~\kev\ in the spectra of BR01 and
\citet{lee01}. It is possible that Fe L may contribute to this
excess.} in the RGS spectra of \mcg\ (BR01). Although it is possible
to ionize Fe to states higher than Fe~\textsc{xviii} and leave sufficient
quantities of O~\textsc{viii} to produce a line, this will result in a
highly ionized atmosphere with a weak \fe\ line and would not increase
the strength of the O line. 

\subsection{Multiple \fe\ line components}
\label{sub:multi}
In order to adequately fit the \fe\ line in the \textit{ASCA} spectra
of \mcg\ with the ionized disc models, it was necessary to add a
separate diskline component to the model. This enabled the model to
fit both the broad red wing of the \fe\ line (through the ionized disc
component) and the height of the profile (through the diskline)
simultaneously. The diskline is too wide to originate from a distant
reflector, such as a molecular torus \citep*[e.g.,][]{kro94}, so it
must arise from elsewhere on the disc. Possible mechanisms for a
second line component include reflection of X-rays from the central
engine as a result of any warping or flaring of the disc
\citep[e.g.][]{bla99}, and reflection from a second illumination event
on the disc which might be expected from a magnetically active and
patchy corona \citep*[e.g.,][]{gal79,haa93}.

In the first scenario, the flared region of the disc would see the
same $\Gamma$ as the inner regions, but the illuminating flux would be
much lower and neutral reflection would dominate. To test this idea,
the diskline component was replaced by a constant density
reflector. This component had its photon-index fixed to the same value
as the inner reflector, and we fit for the value of the ionization
parameter. These results are shown in Table~\ref{table:fits2}.
\begin{table*}
\begin{minipage}{148mm}
\caption{Results of fitting the 1994 \textit{ASCA} data of \mcg\ with
two different ionized reflection spectra. The reflection fraction was
frozen at unity for all models. As before, the inner and outer radii
are reported in gravitational radii, and the inclination angle is in
degrees. In the upper part of the table the second reflector is a
constant density model which had its $\Gamma$ fixed to
be the same as the hydrostatic model. In the
lower region of the table the second reflector is another hydrostatic
model which was drawn from the same grid of models as the first.}
\label{table:fits2}
\begin{tabular}{@{}ccccccccc}
Model & $\Gamma_1$ & $r_{in1}$ & $r_{out1}$ & $\log \xi$/$\Gamma_2$ &
$r_{in2}$ & $r_{out2}$ & $i$ & $\chi^2$/dof \\ \hline 
2xFe ion. disc + 2xFe ion. disc (constant density) & 1.92 & 6.1 & 17 & 1.16 & 38 & 282 & 1.6 & 1665/1649 \\
 & 1.91 & 6.0 & 10 & 3.39 & 22 & 25 & 6.7 & 1656/1649 \\ \hline 
2xFe ion. disc + 2xFe ion. disc & 1.78 & 6.6 & 12 & 2.14 & 18 & 25 & 0.02 & 1672/1649 \\ \hline
\end{tabular}
\end{minipage}
\end{table*}
We find that neutral reflection from a large region of the disc can
account for the second \fe\ line component. However, this result is
not unique, as ionized reflection from a smaller, more central region
of the disc can also fit the data.

If the second \fe\ line component results from another illumination
event then it is not necessary for the $\Gamma$ of the two events to be
the same, and the observed photon-index is a weighted sum of the
two. Models of this type were constructed using two hydrostatic
ionized disc models, and the results are also shown in
Table~\ref{table:fits2}. A decent fit to the 1994 data was found with
the inner reflector subject to a $\Gamma=1.78$ power-law and the outer
to a softer $\Gamma=2.14$ power-law, although a very low value of
the inclination angle was found in this case suggesting further
reflection components are necessary. 

These results emphasize that it is very difficult to fit the 1994 hard
 X-ray spectrum of \mcg\ with a simple ionized disc model. At least
 two different reflection components are needed to adequately fit the
 \fe\ line profile which suggests a very complicated illumination geometry.
 However, multiple \fe\ line components are a natural consequence of
 the magnetic flare model, and could provide the explanation to the
 perplexing variability of the \fe\ line. In contrast, a very
 acceptable fit to the same data was obtained using just one
 reflection component from a predominantly neutral disc which suggests
 a very simple illumination geometry. Unfortunately, current data
 cannot distinguish between the two cases, but our calculations show
 that measurements of soft X-ray disc emission features could lead to
 constraints on the ionization state of the accretion disc surface.

\section{Conclusions}
\label{sect:concl}
The main results of this paper are:
\begin{itemize}
\item The 1994, 1997 and 1999 \textit{ASCA} observations of \mcg\ can
be fit with a relativistic ionized disc model which has twice solar Fe
abundance. However, a diskline component must be included to fully
account for the shape of the \fe\ line.

\item Alternatively, the 1994 data can be well fit by a single
reflection component from a weakly ionized ($\xi = 10$) constant
density disc with twice solar Fe abundance. In this case, one cold
\fe\ line at 6.4~\kev\ can completely describe the observed line.
 
\item Both reflection models predict that the EW of
the O \textsc{viii} K$\alpha$ line is $\sim 10$~eV. This can be made
larger by increasing the O abundance, but O~\textsc{vii} lines
will eventually become prominent. The ionized disc models also predict
that Fe~\textsc{xvii} \& Fe~\textsc{xviii} L$\alpha$ lines will be
found along with the O~\textsc{viii} line.

\item These results cast doubt on the claims of relativistic lines
found in the soft X-rays by BR01 because broad Fe L lines are not seen
in their spectra, and EWs$> 100$~eV are required for the O
\textsc{viii} and C \textsc{vi} lines. However, \mcg\ was in an
unusually low state when observed by BR01, so we cannot strictly rule
out their result.

\item Emission from multiple regions on the disc was needed to
adequately fit the \fe\ line profile with the ionized disc models. The
data are unable to distinguish between secondary reflection from a
flared disc, or from other independent illumination/reflection events.
\end{itemize}

\section*{Acknowledgments}
The authors thank S. Vaughan and R. Morales for valuable discussions,
and K. Iwasawa for assistance with the \textit{ASCA} data.  DRB
acknowledges financial support from the Commonwealth Scholarship and
Fellowship Plan and the Natural Sciences and Engineering Research
Council of Canada. ACF thanks the Royal Society for support.


\bsp 

\label{lastpage}

\end{document}